# Étude de la composition chimique des fluoroalkylamines utilisés en agriculture et en médecine dans le cadre de l'incinération par plasma entre 500 K et 20.000 K

I. Pafadnam, N. Kohio, W.C. Yaguibou, A.K. Kagoné,

Z. Koalaga et P. André

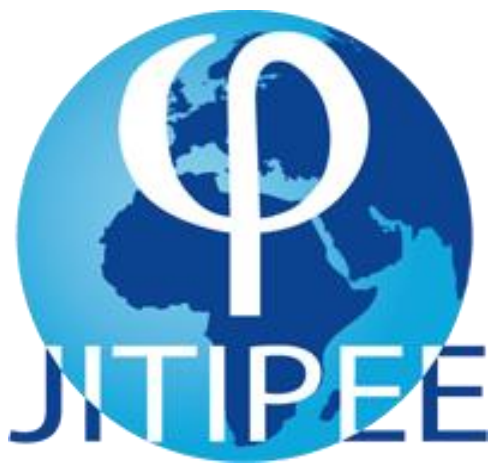

# Étude de la composition chimique des fluoroalkylamines utilisés en agriculture et en médecine dans le cadre de l'incinération par plasma entre 500 K et 20.000 K

Ibrahim Pafadnam[(1)], Nièssan Kohio[(1)], Wêpari Charles Yaguibou[(1)], Abdoul Karim Kagoné[(1)], Zacharie Koalaga[(1)] et Pascal André [(2)]

(1) Université Joseph Ki-ZERBO, Laboratoire des matériaux et environnement (LAME), OUAGADOUGOU, BURKINA FASO
(2) Université Clermont Auvergne, CNRS, Laboratoire de Physique de Clermont, F-63000 CLERMONT-FERRAND, FRANCE
pafadnamib@yahoo.fr

***Résumé*** *- Les principes actifs contenant des composés organiques fluorés comme les fluoroalkylamines ou les molécules à base de pyrimidine sont prometteurs dans le domaine de l'agriculture (pesticides et herbicides) et de la pharmacologie (antibiotiques). L'utilisation massive de ces molécules aura pour conséquence une augmentation massive de déchets contenant ce type de molécules. Les pays développés ont des politiques contraignantes en matière de gestion de déchets ce qui n'est pas le cas dans les pays en voie de développement. Dans ces derniers, nous assistons à une prolifération des aires de stockage et des éliminations à l'air libre des déchets parfois issus des pays développés. Ces pratiques ont d'énormes conséquences sur l'environnement comme la pollution de l'air, des sols et de l'eau et par conséquent sur la santé humaine. L'une des solutions déjà éprouvé sur des déchets solides serait l'utilisation de torches à plasma. Ces torches peuvent atteindre des températures élevées (5.000 K à 20.000 K). Cependant, l'utilisation de ces moyens de traitement n'est pas sans danger puisque des molécules toxiques ou létales pourraient être produites. Afin d'appréhender ces difficultés, nous proposons d'étudier l'influence de l'air sur la composition chimique d'un plasma à base de fluoroalkylamines (trifluoroéthylamine : $C_2H_4F_3N$, nonafluoropentylamine : $C_5H_4F_9N$, …), à la pression atmosphérique et à l'équilibre thermodynamique local (E.T.L), dans une gamme de températures allant de 500 K à 20.000 K. Afin d'obtenir la composition chimique du plasma, nous utilisons la méthode de minimisation de l'énergie libre de Gibbs. Les résultats obtenus montrent que des espèces chimiques gazeuses dangereuses et toxiques comme $CF_2$, CO, HCN et HF apparaissent aux basses températures avec de forte concentration.*

**Mots clés** : *fluoroalkylamines, incinération, plasma, composition chimique, énergie libre de Gibbs*

**Introduction**

L'utilisation des plasmas thermiques notamment au travers de torches à plasma spécifiques et dédiés à la destruction de déchet toxique est prometteuse [1]. De nombreuses études théoriques et expérimentales ont été menées afin de mieux comprendre ces milieux réactifs [2-8]. Ainsi, les torches à plasma ont montré un vaste champ d'applications dans lequel le traitement ou l'élimination des déchets toxiques occupent une place de choix.

De plus, les enjeux environnementaux et climatiques liés au traitement des déchets toxiques ne sont pas seulement d'ordre écologique mais aussi d'ordre sanitaire. En effet, depuis la révolution industrielle au XIX$^e$ siècle, apparaissent de nouveaux types de déchets de plus en plus nombreux. Les niveaux atteints ont obligé les pouvoirs publics à développer de nouvelles filières de gestion des déchets, comme l'incinération industrielle [9]. L'incinération des déchets présente incontestablement l'avantage de réduire le volume des déchets. Toutefois, la question des risques que la technique fait courir à l'Homme et à l'environnement reste posée, notamment en raison des rejets gazeux dans l'atmosphère.

Selon la convention de Bâle (22 mars 1989), les déchets médicaux sont les plus dangereux après ceux qui sont radioactifs. Au Burkina Faso, avec un taux d'occupation de lits des hôpitaux publics de 46,3% en 2015 et 52,3% en 2017 [10], les installations actuelles de traitement de déchets ont du mal à résorber les déchets produits pour de multiples raisons. En 2015, le centre hospitalier universitaire Yalgado Ouédraogo produisait par jour 176 kg de déchets biomédicaux hautement souillés et utilisait 150 sacs poubelles par jour pour la collecte des déchets de tout type [11].

Les crises sanitaires, comme celle de l'épidémie de COVID-19 peuvent conduire à une surproduction de déchets à risques infectieux qui doivent être détruits. De plus, de nombreuses nouvelles molécules basées sur les fluoroalkylAmines (trifluoroethylamine : $C_2H_4F_3N$, nonafluoropentylamine : $C_5H_4F_9N$, …) sont de plus en plus utilisées dans l'agriculture et en pharmacologie notamment sous forme de spray [12]. Les produits non utilisés doivent être détruits lors de l'incinération par plasma, tout en minimisant les rejets atmosphériques de polluants grâce à l'élimination ou à la réduction des substances imbrûlées.

L'objectif final de ce travail est de déterminer l'influence de l'air sur la composition chimique du plasma formé de fluoroalkylamines à la pression atmosphérique et à l'équilibre thermodynamique local (E.T.L) dans une gamme de températures allant de 500 K à 20.000 K. Cette gamme de température couvre les températures mesurées dans les torches à plasma. Dans le cadre de cette étude, nous prenons en compte uniquement la phase gazeuse. Nous considérons que l'air est constitué de 20% de dioxygène et 80% de diazote, et que les différents pourcentages retenus sont des pourcentages volumiques. Les calculs sont menés à la pression atmosphérique. Afin d'étudier des cas tests nous choisissons comme type de fluoroalkylamines, le trifluoroéthylamine ($C_2H_4F_3N$) et le nonafluoropentylamine ($C_5H_4F_9N$).

Dans une première partie, nous décrirons la méthode de calcul permettant d'obtenir la composition chimique à pression atmosphérique. Dans une seconde partie, nous présenterons les résultats obtenus. Finalement, nous conclurons.

**1. Méthode de calcul de la composition chimique.**

Il existe plusieurs méthodes pour la détermination de la composition chimique dans un plasma. A l'équilibre thermodynamique, lorsque l'on veut prendre en compte un grand nombre d'espèces chimiques, deux grandes familles de méthodes sont habituellement utilisées ; l'une est basée sur la loi d'action de masse [2, 13, 14] et l'autre sur la minimisation de l'énergie libre de Gibbs [3, 6, 15]. Dans ce travail nous avons utilisé la méthode de minimisation de l'énergie libre de Gibbs, car elle est très efficace et facile à mettre en œuvre [3]. En outre, elle nécessite la connaissance des données thermodynamiques spécifiques de l'ensemble des espèces chimiques qui constitue le mélange [3, 8]. Cette méthode peut être aussi adaptée aux plasmas à multiples températures [3]. Dans les perspectives de travailler sur les plasmas en déséquilibres thermiques, nous avons jugé nécessaire de ne pas limiter la méthode au cas de l'équilibre thermodynamique local. Les multiplicateurs de Lagrange sont utilisés pour minimiser l'énergie libre de Gibbs (enthalpie libre) ; ce qui entraîne une convergence très rapide. Les phases solides ne sont pas considérées dans le calcul de composition. Leurs prises en compte dans les calculs auraient pour conséquence de minimiser la présence de certaines molécules comme les espèces carbonées dans le cas de la solidification du carbone [6]. Cependant, nous supposons que la torche à plasma sera réglée de telle manière que peu de matières issues des phases condensées soient produites. Les résultats en phase gazeuse de la composition chimique, présentés dans cet article, seront alors proches des gaz obtenus durant le fonctionnement de la torche à plasma.

**2. Mélanges considérés.**

Dans cette étude nous considérons que l'air est constitué de 80% de diazote ($N_2$) et 20% de dioxygène ($O_2$), les autres constituants (*Ar*, $CO_2$, $CH_4$, …) ont été totalement négligés. L'air contient donc des atomes d'azote (*N*) et d'oxygène (*O*). Les fluoroalkylamines étant constitués des atomes de carbone (*C*), de fluor (*F*), d'hydrogène (*H*) et d'azote (*N*), le mélange fluoroalkylamine-air est donc constitué uniquement des atomes *C*, *F*, *H*, *N* et *O*, permettant d'obtenir des plasmas de type $C_xF_yH_zN_tO_u$. Les indices *x*, *y*, *z*, *t* et *u* sont des variables entières ou fractionnaires fonction des proportions du mélange.

Les différents pourcentages retenus sont des pourcentages volumiques. Ces variables sont déterminées à partir des pourcentages d'air et de fluoroalkylamines du mélange.
Considérons le mélange trifluoroéthylamine-air, avec *V* le volume du mélange, $P_1$ le pourcentage d'air dans le mélange et $P_2$ le pourcentage volumique de trifluoroéthylamine dans le mélange.

- le volume d'air dans le mélange est : $V_{air} = P_1 * V$     (1)
- le volume de trifluoréthylamine dans le mélange est : $V_{trif} = P_2 * V$     (2)
- le nombre d'atomes de carbone est : $N_1 = \frac{2P_2*V}{V_m} * \mathcal{N}$     (3)
- le nombre d'atomes de fluor est : $N_2 = \frac{3P_2*V}{V_m} * \mathcal{N}$     (4)
- le nombre d'atomes d'hydrogène est : $N_3 = \frac{4P_2*V}{V_m} * \mathcal{N}$     (5)
- le nombre d'atomes d'azote est : $N_4 = \frac{V\left(P_2 + \frac{8}{5}*P_1\right)}{V_m} * \mathcal{N}$     (6)
- le nombre d'atomes d'oxygène est : $N_5 = \frac{2P_1*V}{5V_m} * \mathcal{N}$     (7)
- le nombre total d'atomes est : $N_t = (2P_1 + 10P_2)\frac{V}{V_m} * \mathcal{N}$     (8)

Dans ces relations, $\mathcal{N}$ est le nombre d'Avogadro et $V_m$ est le volume molaire.

*JITIPEE vol.8:n°1: 1 (2023)*Nous déduisons que :

- $x = \frac{2P_2}{2P_1+10P_2}$ (9)

- $y = \frac{3P_2}{2P_1+10P_2}$ (10)

- $z = \frac{4P_2}{2P_1+10P_2}$ (11)

- $t = \frac{8P_1+5P_2}{5(2P_1+10P_2)}$ (12)

- $u = \frac{2P_1}{5(2P_1+10P_2)}$ (13)

A partir de ces relations, nous obtenons les pourcentages des éléments de base *C, F, H, N, O* pour les autres mélanges considérés. Ces pourcentages sont donnés en fonction du mélange ou du gaz considéré dans le Tableau 1.

| Mélanges | Notation | x | y | z | t | u |
|---|---|---|---|---|---|---|
| *100% d'air sec* | **air** | 0,00 | 0,00 | 0,00 | 80,00 | 20,00 |
| *100% trifluoroéthylamine* | **trif** | 20,00 | 30,00 | 40,00 | 10,00 | 00,00 |
| *100% nonafluoropentylamine* | **nonaf** | 26,32 | 47,37 | 21,05 | 05,26 | 00,00 |
| *99%trif + 1%air* | **mel1** | 19,96 | 29,94 | 39,92 | 10,14 | 00,04 |
| *50%trif + 50%air* | **mel50** | 16,67 | 25,00 | 33,33 | 21,67 | 03,33 |
| *99%nonaf + 1%air* | **mel'1** | 26,29 | 47,32 | 21,03 | 05,34 | 00,04 |
| *50%nonaf + 50%air* | **mel'50** | 23,81 | 42,86 | 19,05 | 12,38 | 03,33 |

**Tableau 1**. Notation et valeurs des variables x, y, z, t et u des différents plasmas étudiés. (trif=trifluoroéthylamine, nonaf= nonafluoropentylamine).

### 3. Calcul de la composition chimique.

Les espèces chimiques considérées dans le plasma de mélange fluoroalkylamine-air sont :

- les électrons : $e^-$ ;
- les espèces monoatomiques (18): *C, $C^+$, $C^{2+}$, $C^-$, F, $F^+$, $F^-$, H, $H^+$, $H^-$, N, $N^+$, $N^{2+}$, $N^-$, O, $O^+$, $O^{2+}$, $O^-$* ;
- les espèces diatomiques (27): *$C_2$, $F_2$, $H_2$, $N_2$, $O_2$, CF, CH, CN, NF, NH, CO, NO, OH, HF, $C_2^+$, $H_2^+$, $N_2^+$, $O_2^+$, $NO^+$, $CF^+$, $CH^+$, $CN^+$, $NH^+$, $CO^+$, $C_2^-$, $H_2^-$, $CN^-$* ;
- les espèces polyatomiques (13): *$CO_2$, $NO_2$, $NO_3$, $CF_2$, $CF_3$, $CH_4$, $C_2F$, $C_2F_2$, $C_2F_3$, $NH_3$, $N_2H_4$, HCN, $NO_2^-$*.

Pour déterminer la composition à l'équilibre, il est au préalable nécessaire de connaitre les potentiels chimiques spécifiques de l'ensemble des espèces chimiques présentes dans le plasma [16]. Pour les électrons, les espèces atomiques et moléculaires, leur potentiel chimique peut être déterminé en utilisant les données de l'enthalpie spécifique et l'entropie spécifique tabulées par les tables de JANAF, NIST [17], McBride et al [18], Bendjebbar et al [4]. Nous utilisons essentiellement les données thermodynamiques lissées et tabulées par McBride et al [18] pour le calcul des propriétés thermodynamiques spécifiques.

*1-4*

La capacité calorifique spécifique, l'enthalpie spécifique et l'entropie spécifique sont obtenues en fonction de la température et de la gamme de température par les relations suivantes [18] :

$$\frac{C_p^0(T)}{R} = a_1 T^{-2} + a_2 T^{-1} + a_3 + a_4 T + a_5 T^2 + a_6 T^3 + a_7 T^4 \quad (14)$$

$$\frac{h^0(T)}{RT} = -a_1 T^{-2} + \frac{a_2 \ln(T)}{T} + a_3 + a_4 \frac{T}{2} + a_5 \frac{T^2}{3} + a_6 \frac{T^3}{4} + a_7 \frac{T^4}{5} + \frac{b_1}{T} \quad (15)$$

$$\frac{s^0(T)}{R} = -\frac{a_1 T^{-2}}{2} - a_2 T^{-1} + a_3 \ln(T) + a_4 T + a_5 \frac{T^2}{2} + a_6 \frac{T^3}{3} + a_7 \frac{T^4}{4} + b_2 \quad (16)$$

où $R$ est la constante des gaz parfaits, $T$ la température en Kelvins. Les coefficients $a_i$ et $b_i$ pour chaque espèce chimique sont tirés des tables de la JANAF [4, 17, 18]. Les potentiels chimiques spécifiques de chaque espèce chimique sont directement obtenus, pour chaque température $T$ donnée par [3, 4] :

$$\mu^0 = h^0 - Ts^0 \quad (17)$$

Pour déterminer les concentrations des différentes espèces chimiques nous utilisons la méthode de minimisation de l'énergie libre de Gibbs du plasma étudié [3, 19-23]. Les concentrations des espèces chimiques symbolisées par un point $Y(y_1, y_2, y_3, \ldots, y_M)$ doivent satisfaire la neutralité électrique et la conservation du nombre de noyaux dans le plasma. Ces deux conditions peuvent s'écrire :

$$\sum_{i=1}^{M} a_{ij} y_j = b_j \qquad (j = 1, 2, 3, \ldots, m) \quad (18)$$

où $M$ est le nombre d'espèces chimiques dans le mélange, $m$ est le nombre de types de noyaux différents (électrons compris), $a_{ij}$ est le nombre de noyaux de type $j$ contenus dans l'espèce chimique $i$, $b_j$ représente le nombre initial de noyaux de type $j$. Dans nos plasmas, les noyaux de types différents sont au nombre de six (6) : $e^-$, C, F, H, N, O.

L'énergie libre de Gibbs établie pour les plasmas multi-températures est donnée par [3, 7] :

$$G = \sum_{i=1}^{M} n_i \left[ \mu_i^0 + RT_i \ln\left(\frac{n_i T_i}{\sum_{i=1}^{M} n_i T_i}\right) + RT_i \ln\left(\frac{P}{P^0}\right) \right] \quad (19)$$

où $\mu_i^0$ est le potentiel chimique spécifique de l'espèce $i$, $T_i$ la température de l'espèce chimique $i$, $P_0$ est la pression de référence, $P$ la pression totale dans le plasma, $n_i$ la densité numérique de l'espèce chimique $i$. Les seules inconnues de la relation (19) sont les potentiels chimiques spécifiques $\mu_i^0$ et les densités numériques $n_i$ des différentes espèces chimiques pour une pression et une température fixées.

L'énergie libre de Gibbs calculée au point $Y(y_1, y_2, y_3, \ldots, y_M)$ donne :

$$G(Y) = \sum_{i=1}^{M} y_i \left( C_i + RT_i \ln \frac{y_i T_i}{\sum_{k=1}^{M} y_k T_k} \right) \quad (20)$$

avec

$$C_i = \mu_i^0 + RT_i \ln \frac{P}{P^0} \quad (21)$$

Les valeurs $y_i$ sont proportionnelles aux densités $n_i$. Nous devons rechercher le point $Y(y_1, y_2, y_3, \ldots, y_M)$ qui minimise la fonction $G(Y)$ et pour lequel les coordonnées $y_i$ satisfont les conditions suivantes :
- les concentrations $n_i$ doivent être positives d'où $y_i \geq 0 \qquad \forall\, i$ ;
- les coordonnées $y_i$ doivent satisfaire la conservation du nombre de noyaux et la neutralité électrique.

Un développement en série de Taylor d'ordre deux (2) autour du point $Y(y_1, y_2, y_3, \ldots, y_M)$ donne :

$$Q(X) = G(Y) + \sum_{i=1}^{M} \left(\frac{\partial G}{\partial x_i}\right)_{X=Y} (x_i - y_i) + \frac{1}{2}\sum_{i=1}^{M}\sum_{k=1}^{M} \left(\frac{\partial^2 G}{\partial x_i \partial y_k}\right)_{X=Y} (x_i - y_i) \quad (22)$$

Pour prendre en compte les conditions physiques (18), on introduit les multiplicateurs de Lagrange $\pi_j$ et on obtient ainsi la nouvelle fonction $\zeta(X)$ :

$$\zeta(X) = Q(X) + \sum_j \pi_j \left(-\sum_i a_{ij} x_i + b_j\right) \quad (23)$$

$G(X)$ est alors minimale lorsqu'on a [3, 8] :

$$\frac{\partial \zeta(X)}{\partial x_i} = 0 \quad (24)$$

Ce qui conduit à :

$$\frac{f_i}{y_i} + RT_i \left[\frac{x_i}{y_i} - \frac{\sum_k T_k x_k}{\sum_k T_k y_k}\right] - \sum_{j=1}^{m} \pi_j a_{ij} = 0 \quad (25)$$

avec

$$f_i = y_i \left(\mu_i^0 + RT_i \ln \frac{P}{P^0} + RT \ln \frac{y_i T_i}{\sum_k y_k T_i}\right) \quad (26)$$

En utilisant la méthode de Newton-Raphson, on obtient à l'équilibre thermique le système d'équation suivant [4, 24] :

$$\begin{pmatrix} \frac{RT}{n_1} & \ldots & 0 & a_{1,0} & \ldots & a_{1,5} \\ \ldots & \ldots & \ldots & \ldots & \ldots & \ldots \\ 0 & \ldots & \frac{RT}{n_M} & a_{M,0} & \ldots & a_{M,5} \\ a_{1,0} & \ldots & a_{M,0} & 0 & \ldots & 0 \\ \ldots & \ldots & \ldots & \ldots & \ldots & \ldots \\ a_{1,5} & \ldots & a_{M,5} & 0 & \ldots & 0 \end{pmatrix} \begin{pmatrix} \Delta n_1 \\ .. \\ \Delta n_M \\ \Delta \pi_0 \\ \ldots \\ \Delta \pi_5 \end{pmatrix} = \begin{pmatrix} -\mu_1^0 - RT \ln \frac{n_1}{\sum_i^M n_i} - RT \ln \frac{P}{P^0} - \sum_{j=0}^{5} \pi_j a_{i,j} \\ \ldots \\ -\mu_M^0 - RT \ln \frac{n_M}{\sum_i^M n_i} - RT \ln \frac{P}{P^0} - \sum_{j=0}^{5} \pi_j a_{i,j} \\ -\sum_{i=1}^{M} \pi_j a_{i,0} + b_0 \\ \ldots \\ -\sum_{i=1}^{M} \pi_j a_{i,5} + b_5 \end{pmatrix} \quad (27)$$

avec $x_i = n_i$. Dans le cas de notre étude, nous avons pris en compte cinquante-neuf (59) espèces chimiques donc $M = 59$. Le système d'équations (27) est donc un système de soixante-cinq équations à soixante-cinq inconnues, qui sont les nombres de moles des cinquante-neuf espèces chimiques plus les six multiplicateurs de Lagrange.

La méthode numérique consiste à affecter initialement et arbitrairement des valeurs aux nombres de moles $n_i$ avec $i \in [1, 59]$ et aux multiplicateurs de Lagrange $\pi_j$ avec $j \in [0, 5]$. Les valeurs des nombres de moles doivent vérifier la condition (19) c'est-à-dire :

$$\begin{cases} n_i \geq 0 & \forall\, i \in [1, 59] \\ \sum_{i=1}^{N} a_{ij} y_j = b_j & \forall\, j \in [0, 5] \end{cases} \quad (28)$$

On résout le système d'équations (27) et les nouvelles valeurs des nombres de moles et des multiplicateurs de Lagrange sont obtenues par la relation suivante :

$$\begin{cases} n_i = n_i + \lambda \Delta n_i & \forall\, i \in [1,\ 59] \\ \pi_j = \pi_j + \lambda \Delta \pi_j & \forall\, j \in [0,\ 5] \end{cases} \quad (29)$$

Le paramètre de correction $\lambda$ est la plus grande valeur comprise entre 0 et 1 satisfaisant la condition suivante :

$$n_i = n_i + \lambda \Delta n_i > 0 \qquad \forall\, i \in [1,\ 59] \quad (30)$$

Cette étape permet d'éviter les valeurs négatives des nombres de moles qui peuvent apparaître lorsqu'on s'éloigne de la solution. Les nouvelles valeurs des nombres de moles et des multiplicateurs de Lagrange sont utilisées pour un nouveau cycle de calcul. Le critère d'interruption des itérations est fixé par la condition suivante :

$$\Delta n_i < 10^{-15} n_i \quad \forall\, i \in [1,\ 59] \quad (31)$$

## 4. Résultats et discussion.

Dans cette partie, nous présentons les résultats de calcul de composition d'équilibre des différents plasmas étudiés. Pour valider notre programme de calcul, nous avons comparé les résultats de calculs des fractions molaires d'un plasma d'air pur avec les données de Yaguibou [23]. Les écarts observés sont donnés dans le Tableau 2 où sont données les fractions molaires de quelques espèces chimiques majoritaires à savoir les espèces chimiques $e^-$, $N$, $O$, $N_2$, $O_2$, $O^+$ et $O_2^+$ à la pression atmosphérique pour deux valeurs de la température (5.000 K et 15.000 K). Nous remarquons qu'à basse température ($T \leq 5.000\ K$) nos résultats sont en bon accord avec ceux de l'auteur car les écarts relatifs observés varient entre 0% et 1,5%. À haute température ($T \geq 15.000\ K$) la comparaison de nos résultats donne des écarts compris entre 0% et 18%. Les différents écarts peuvent s'expliquer par les données utilisées dans notre programme de calcul. En effet, les données spécifiques ne sont pas calculées, à partir des fonctions de partition, pour chaque espèce chimique et à chaque pas de température, nous utilisons des données lissées issues de différentes sources ce qui induit des résultats différents en fonction des sources utilisées.

| Particules | Yaguibou [23] | Nos résultats | Ecarts | Yaguibou [23] | Nos résultats | Ecart |
|---|---|---|---|---|---|---|
| | **5.000 K** | **5.000 K** | **%** | **15.000 K** | **15.000 K** | **%** |
| $e^-$ | $4{,}16\ 10^{-5}$ | $4{,}16\ 10^{-5}$ | 0 | $3{,}41\ 10^{-1}$ | $3{,}26\ 10^{-1}$ | 4,40 |
| $N$ | $2{,}66\ 10^{-2}$ | $2{,}66\ 10^{-2}$ | 0 | $2{,}39\ 10^{-1}$ | $2{,}63\ 10^{-1}$ | 10,04 |
| $O$ | $3{,}11\ 10^{-1}$ | $3{,}10\ 10^{-1}$ | 0 | $7{,}78\ 10^{-2}$ | $8{,}39\ 10^{-2}$ | 7,84 |
| $N_2$ | $6{,}43\ 10^{-1}$ | $6{,}43\ 10^{-1}$ | 0 | $4{,}12\ 10^{-6}$ | $4{,}61\ 10^{-6}$ | 10,62 |
| $O_2$ | $1{,}97\ 10^{-3}$ | $2{,}00\ 10^{-3}$ | 1,5 | $2{,}80\ 10^{-8}$ | $3{,}27\ 10^{-8}$ | 14,37 |
| $O^+$ | $7{,}48\ 10^{-8}$ | $7{,}48\ 10^{-8}$ | 0 | $5{,}40\ 10^{-2}$ | $5{,}08\ 10^{-2}$ | 5,93 |
| $O_2^+$ | $3{,}32\ 10^{-8}$ | $3{,}32\ 10^{-8}$ | 0 | $1{,}27\ 10^{-7}$ | $1{,}55\ 10^{-7}$ | 18,06 |

**Tableau 2.** Comparaison de nos résultats avec des données de la littérature.

**4.1. Composition d'équilibre du plasma d'air.**

La figure 1 présente la composition d'équilibre du plasma d'air (80% N₂ et 20% O₂) à l'équilibre thermodynamique local et à la pression atmosphérique (1 bar). Sur cette figure, les courbes des densités numériques des espèces chimiques présentes dans le plasma sont représentées en fonction de la température. En examinant ces différents résultats, nous constatons une évolution des densités numériques des différentes espèces chimiques en trois phases :

- la première phase correspond au domaine de température inférieure à 3.000 K ($T < 3.000\ K$). Dans ce cas, les espèces chimiques majoritaires sont les molécules $N_2$, $O_2$ qui sont les molécules de base de l'air et $NO$ ;
- la deuxième phase concerne la plage de températures comprises entre 3.000 K et 8.200 K ($3.000\ K < T < 8.200\ K$). Dans cet intervalle, les espèces chimiques majoritaires sont les espèces neutres N, O qui proviennent des réactions de dissociation des molécules $N_2$, $O_2$ et $NO$. Les ions $N^+$, $O^+$ et $N_2^+$ apparaissent progressivement. La neutralité électrique est rigoureusement établie entre les électrons et les ions $NO^+$ sur cette gamme de température ;
- la troisième partie correspond aux températures supérieures à 8.200 K ($T > 8.200\ K$). Dans cette phase, les espèces neutres disparaissent progressivement ainsi que l'ion $NO^+$. L'électroneutralité s'établit progressivement entre les électrons et les ions $N^+$, $O^+$. La contribution à la production des électrons est principalement due à l'ionisation de l'atome d'azote (*N*). On note aussi l'apparition progressive des ions $N^{++}$ et $O^{++}$ pour les hautes températures.

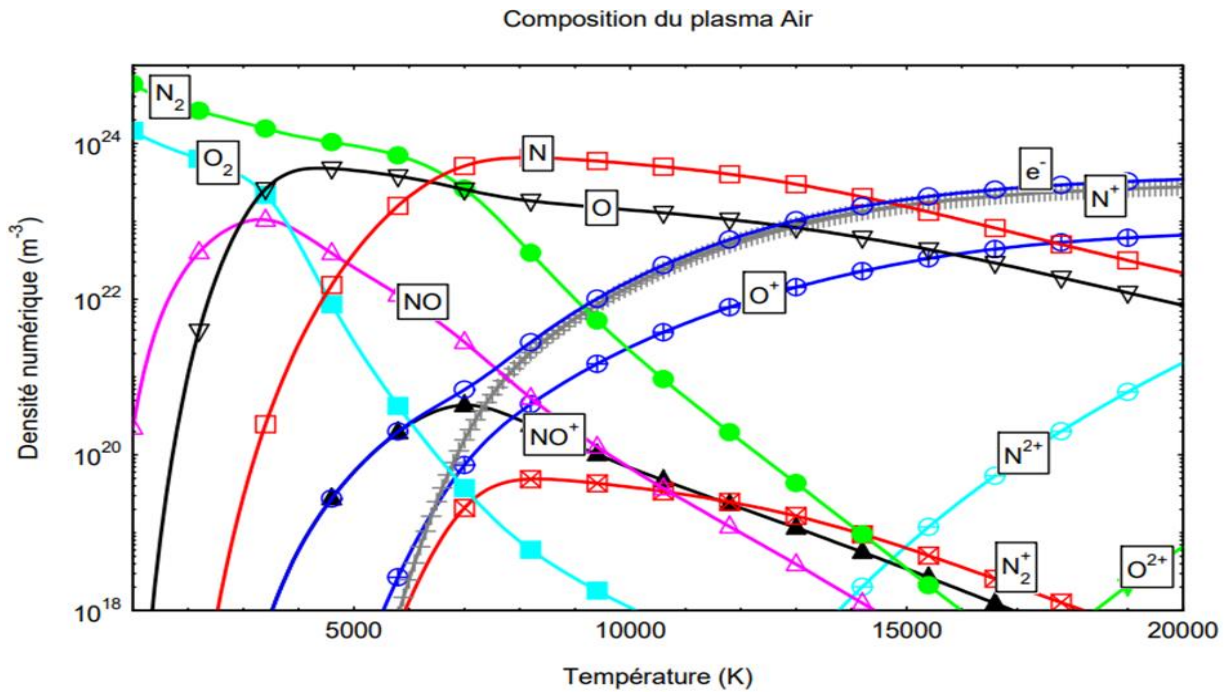

**Figure 1 :** Évolutions en fonction de la température des densités numériques des espèces chimiques du plasma d'air à la pression atmosphérique.

### 4.2. Composition d'équilibre des plasmas de trifluoroéthylamine et de tonafluoropentylamine.

Les figures 2 et 3 présentent les compositions d'équilibre des plasmas de trifluoroéthylamine et de nonafluoropentylamine à l'équilibre thermodynamique local et à la pression atmosphérique (1 bar). Sur ces figures, les densités numériques des espèces chimiques présentes dans les plasmas sont représentées en fonction de la température. En examinant ces différents résultats, nous constatons que :

- pour des températures comprises entre 500 K et 7.500 K, les molécules ($C_4N_2, CH_4, C_2H_4, C_2H_2, C_2F_2, HCN, CH_3, CH_2, C_2F, C_2H, C_3, CF_2$) se dissocient totalement, tandis que les molécules ($C_2, H_2, N_2, CF, CH, CN, HF, NF$) disparaissent progressivement. Ce qui explique les densités croissantes des espèces chimiques telles que le fluor (F), l'hydrogène (H), l'azote (N), le carbone (C) ;
- pour des températures comprises entre 7.500 K et 15.000 K, toutes les molécules ($C_2, H_2, N_2, CF, CH, CN, HF, NF$), se dissocient totalement. Les densités numériques des atomes ($F, H, N, C$) diminuent progressivement, tandis que les espèces ($H^+, N^+, F^+, C^+, CF^+$) apparaissent progressivement ;
- pour des températures comprises entre 15.000 K et 20.000 K, les concentrations des éléments ($F, H, N, C$) continuent de décroitre, tandis que celles des ions ($H^+, N^+, F^+, C^+$) augmentent. Les ions ($CF^+$) disparaissent totalement. On note l'apparition progressive des ions deux fois ionisés ($N^{2+}, C^{2+}$).

### 4.3. Composition d'équilibre des plasmas de trifluoroéthylamine et de nonafluoropentylamine avec de l'air.

Nous présentons dans cette partie, les résultats des calculs de composition d'équilibre du plasma de mélange fluoroalkylamines-air afin de mettre en évidence l'influence de l'air sur un plasma de fluoroalkylamines. Les figures 4 à 7 présentent l'évolution en fonction de la température des densités numériques de certaines espèces chimiques importantes du plasma pour différentes valeurs du pourcentage d'air dans le mélange, à l'équilibre thermodynamique local (E.T.L) et à la pression atmosphérique. On constate que les densités numériques des plasmas de mélange évoluent de manière similaire. En examinant les différents résultats, il ressort que :

- pour le domaine de températures inférieures à 8.000 K, les molécules ($CF_2, HCN, C_2F_2, C_2F$) se dissocient totalement, tandis que les espèces chimiques ($CH, CO, C_2, CF, H_2, N_2, CN, HF, NO$) se dissocient progressivement. Ces dissociations provoquent l'augmentation des densités numériques des espèces chimiques telles que le carbone ($C$), le fluor ($F$), l'hydrogène ($H$), l'azote ($N$) et l'oxygène ($O$) ;
- pour la plage de températures comprises entre 8.000 K et 15.000 K, toutes les espèces moléculaires ($CH, CO, C_2, CF, H_2, N_2, CN, HF, NO$) disparaissent totalement, tandis que les densités numériques des espèces atomiques ($F, H, N, C, O$) diminuent progressivement. Ce qui explique l'apparition des électrons et des espèces ioniques ($H^+, N^+, F^+, C^+, CF^+$). Aussi, on note l'établissement de la neutralité électrique entre les électrons et les ions $C^+$ ;
- pour l'intervalle de températures comprises entre 15.000 K et 20.000 K, les densités numériques des espèces atomiques continuent de décroitre, tandis que celles des ions monoatomiques ($H^+, N^+, F^+, C^+, O^+$) augmentent. On note également l'apparition progressive des ions deux fois ionisés ($N^{2+}, C^{2+}$).

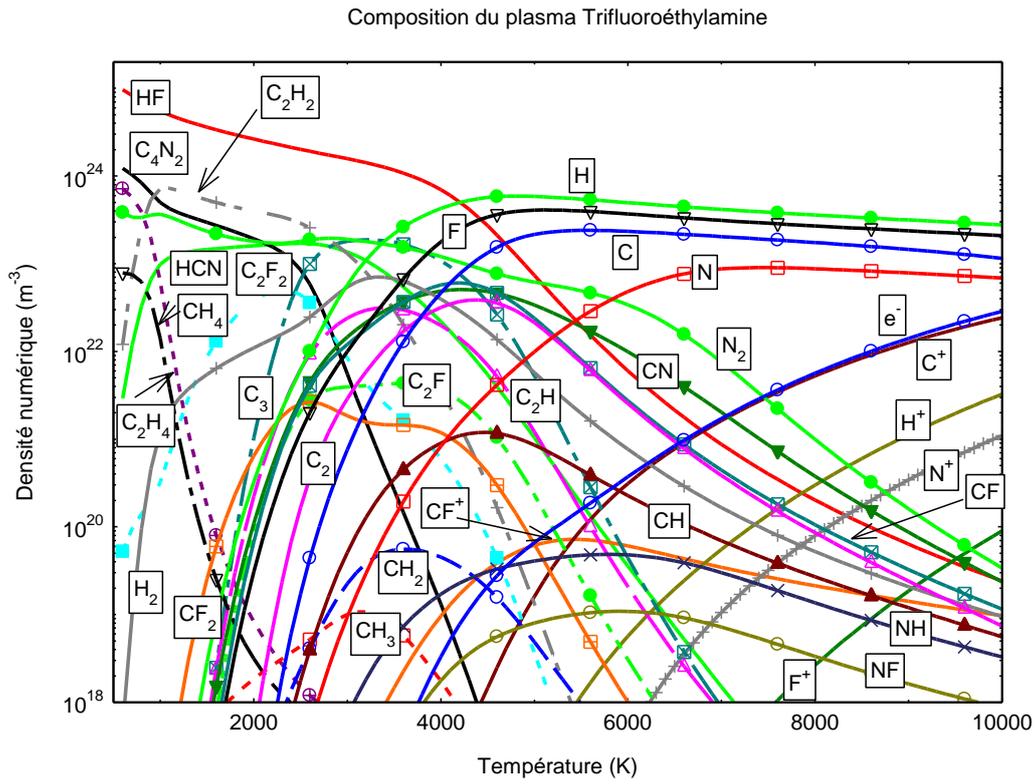

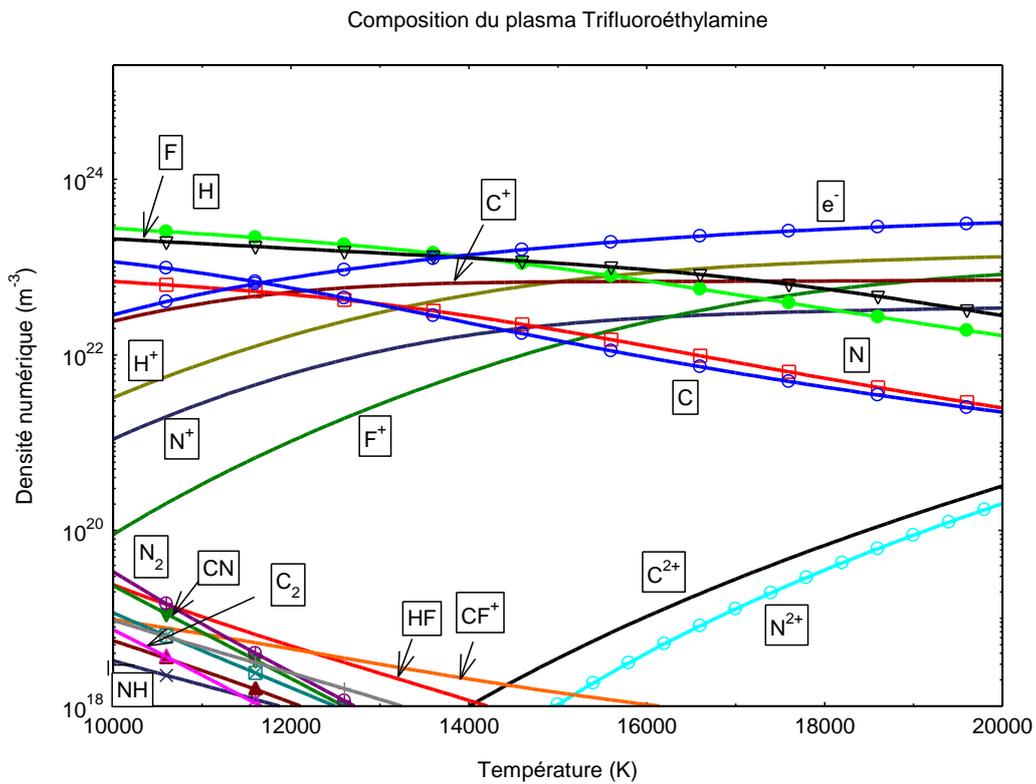

**Figure 2 :** Évolutions en fonction de la température des densités numériques des espèces chimiques du plasma de trifluoroéthylamine à la pression atmosphérique.

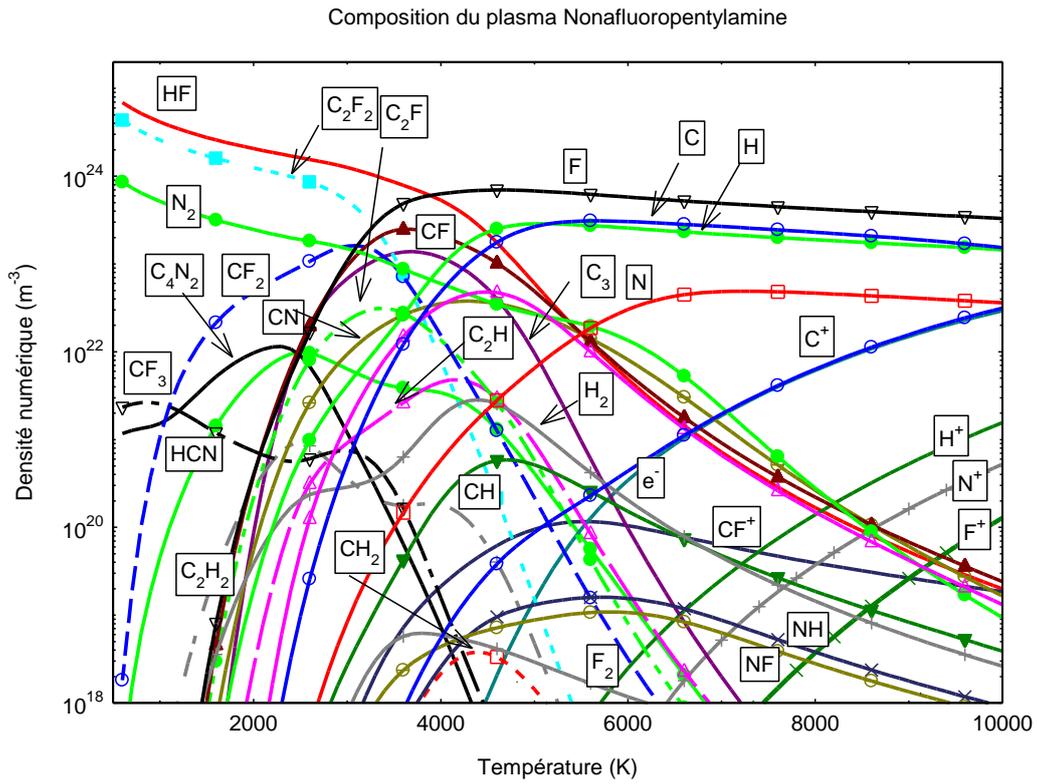

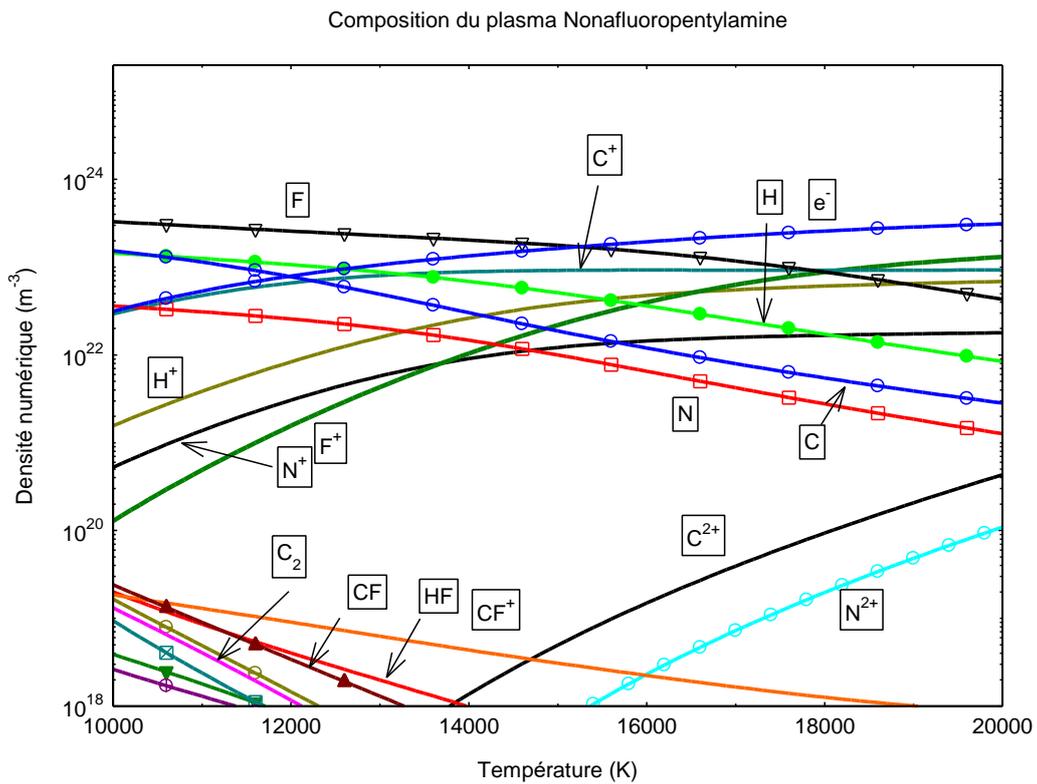

**Figure 3 :** Évolutions en fonction de la température des densités numériques des espèces chimiques du plasma de nonafluoropentylamine à la pression atmosphérique.

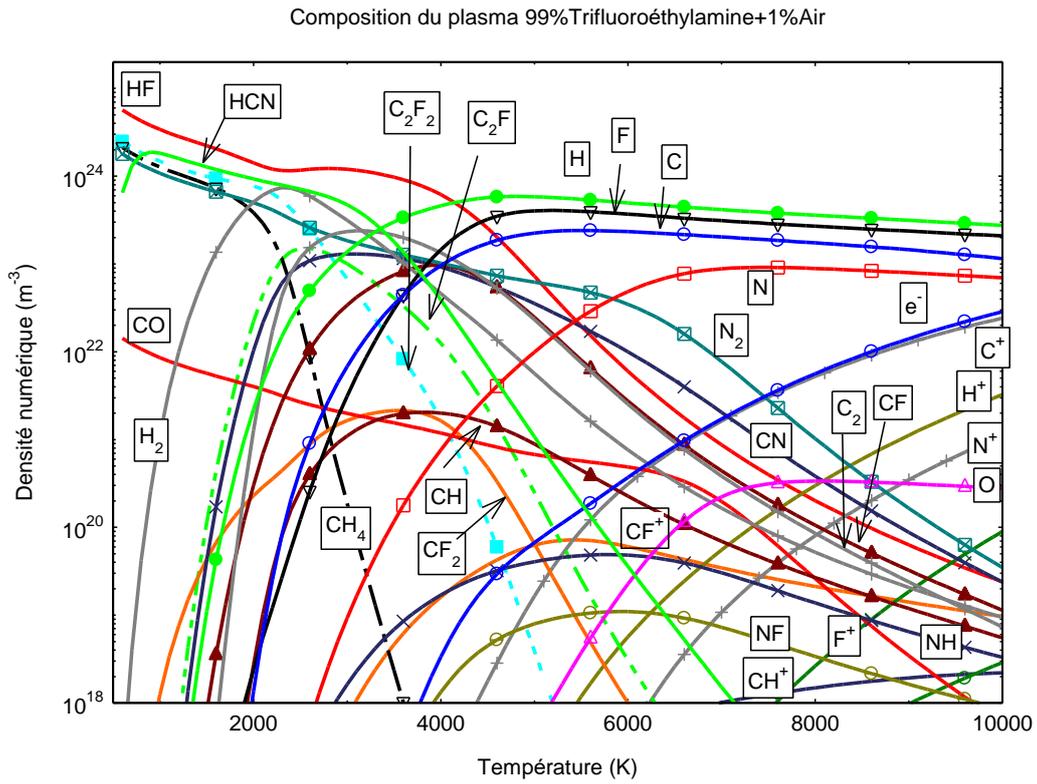

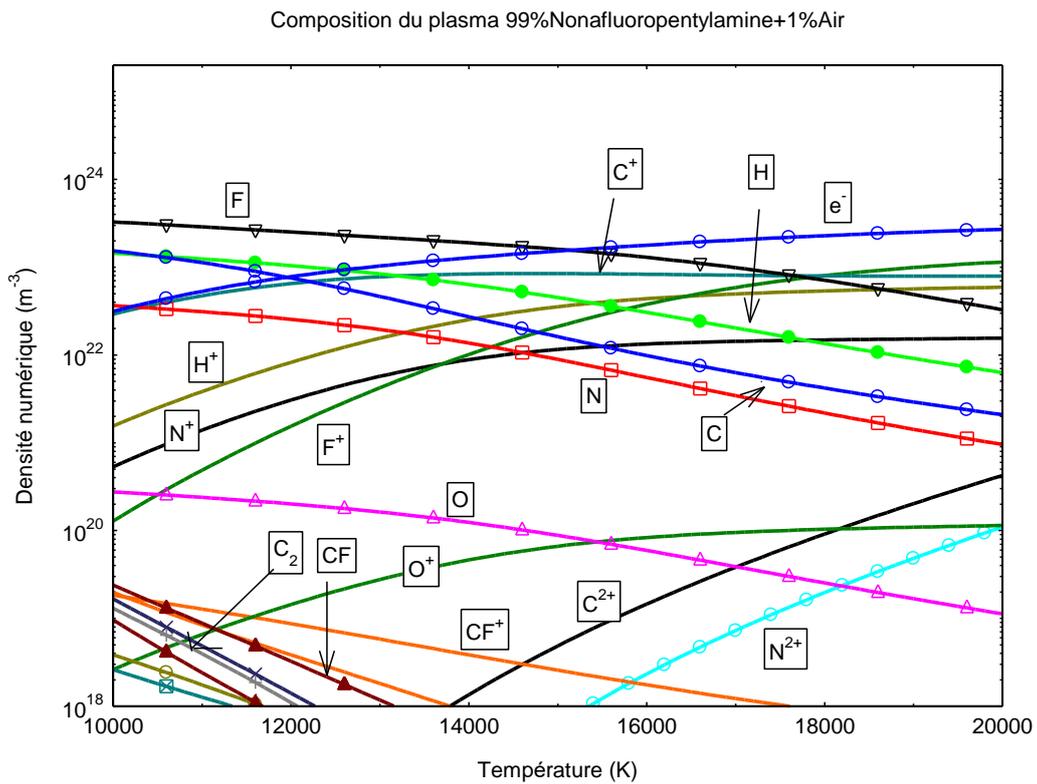

**Figure 4 :** Évolutions en fonction de la température des densités numériques des espèces chimiques du plasma composé de 99% de trifluoroéthylamine et 1% d'air en volume à la pression atmosphérique.

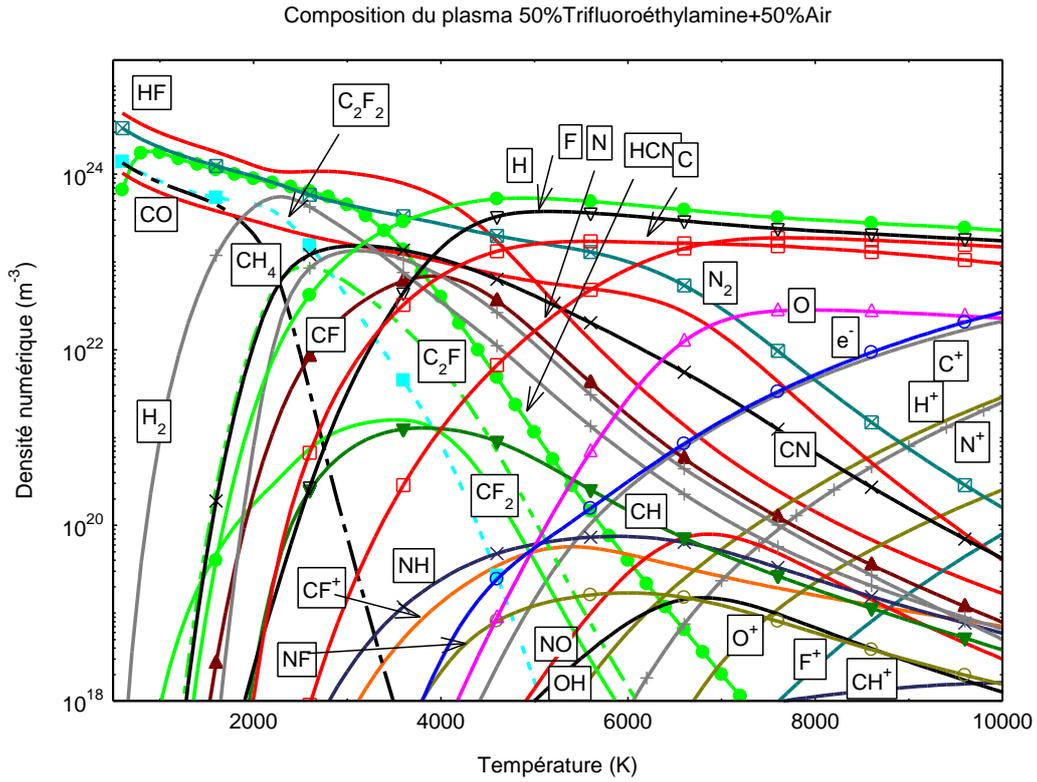

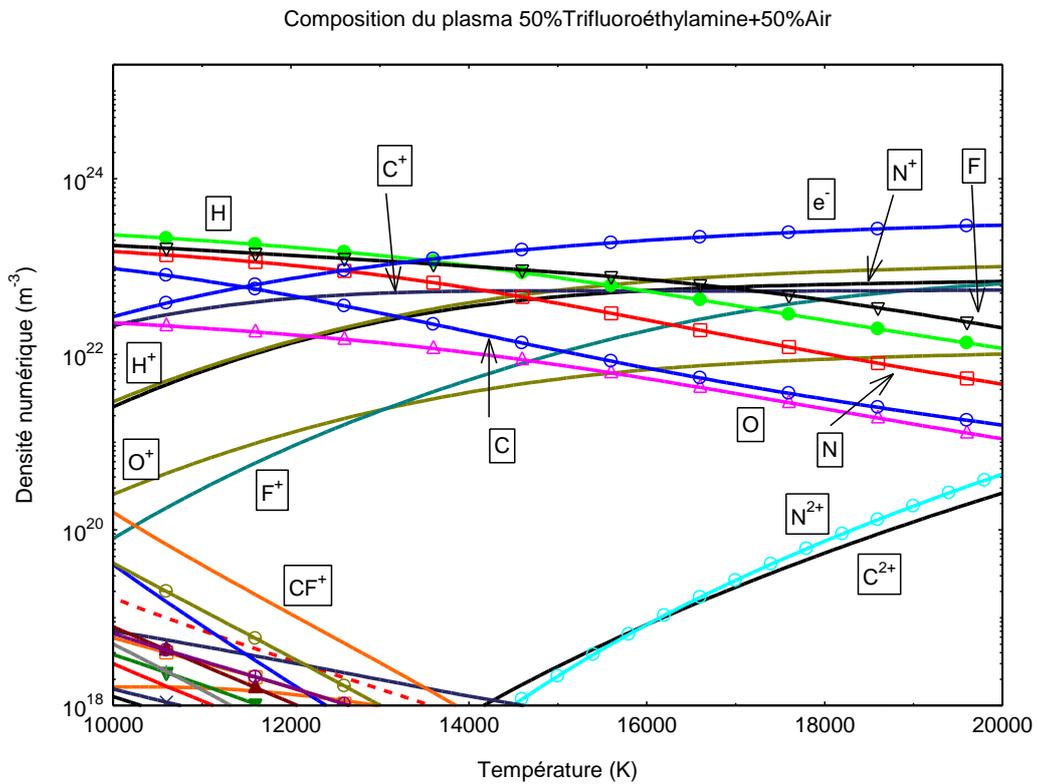

**Figure 5 :** Évolutions en fonction de la température des densités numériques des espèces chimiques du plasma composé de 50% de trifluoroéthylamine et 50% d'air en volume à la pression atmosphérique.

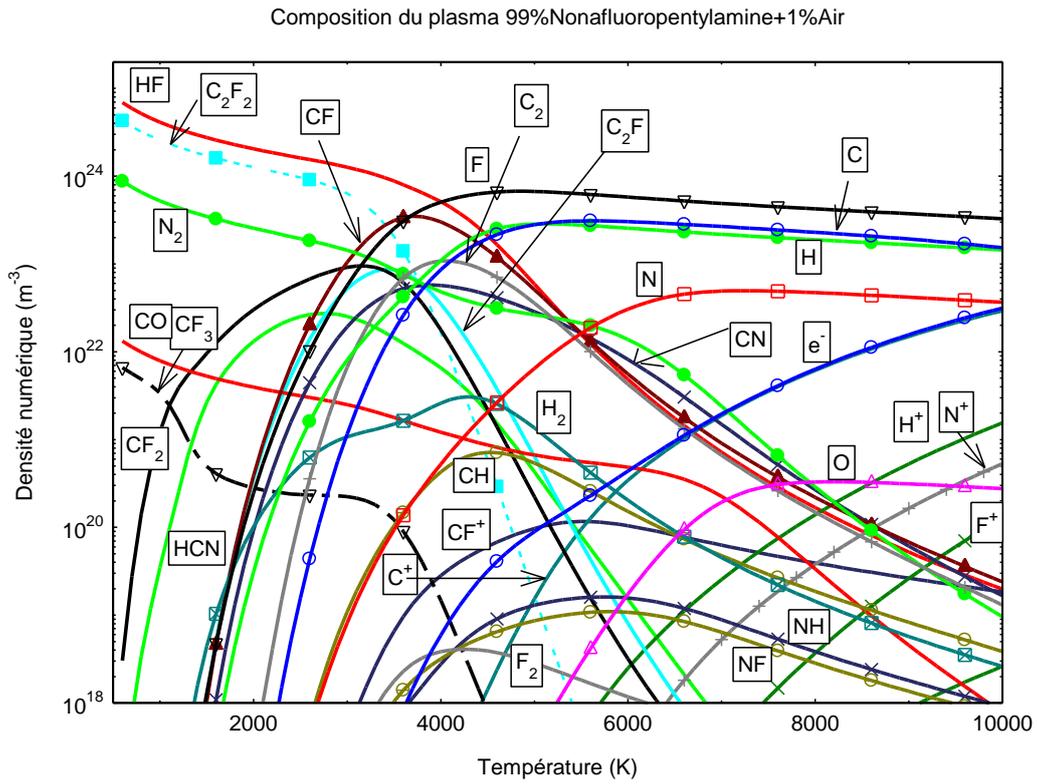

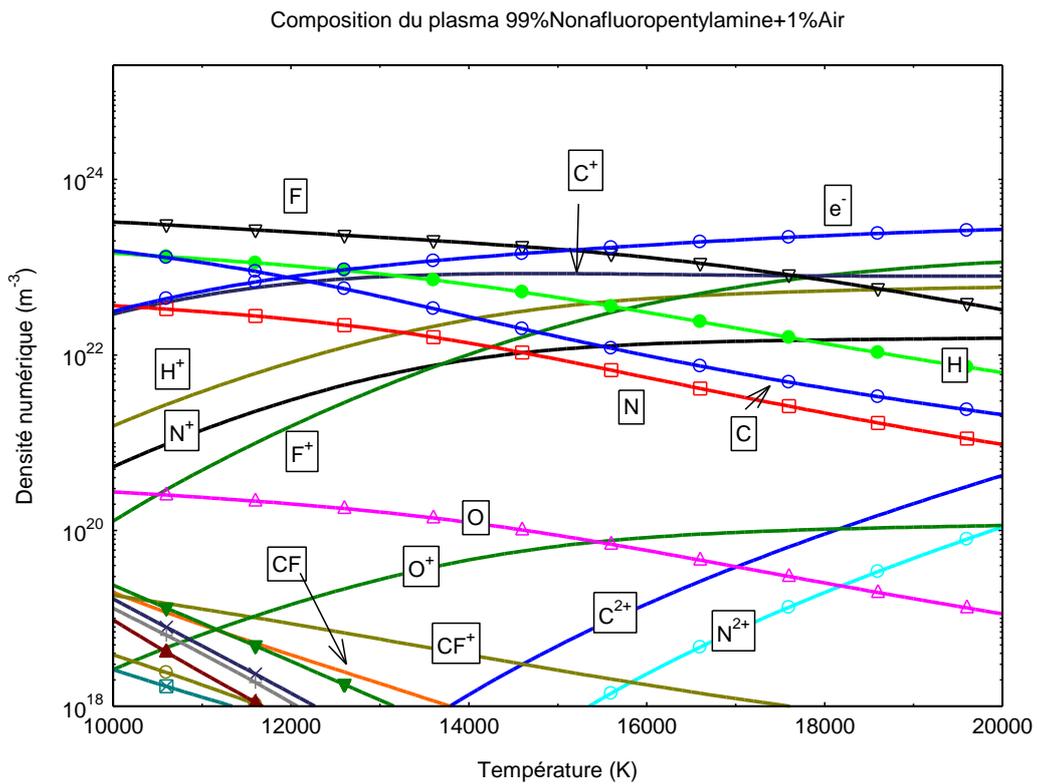

**Figure 6 :** Évolutions en fonction de la température des densités numériques des espèces chimiques du plasma composé de 99% de nonafluoropentylamine et 1% d'air en volume à la pression atmosphérique.

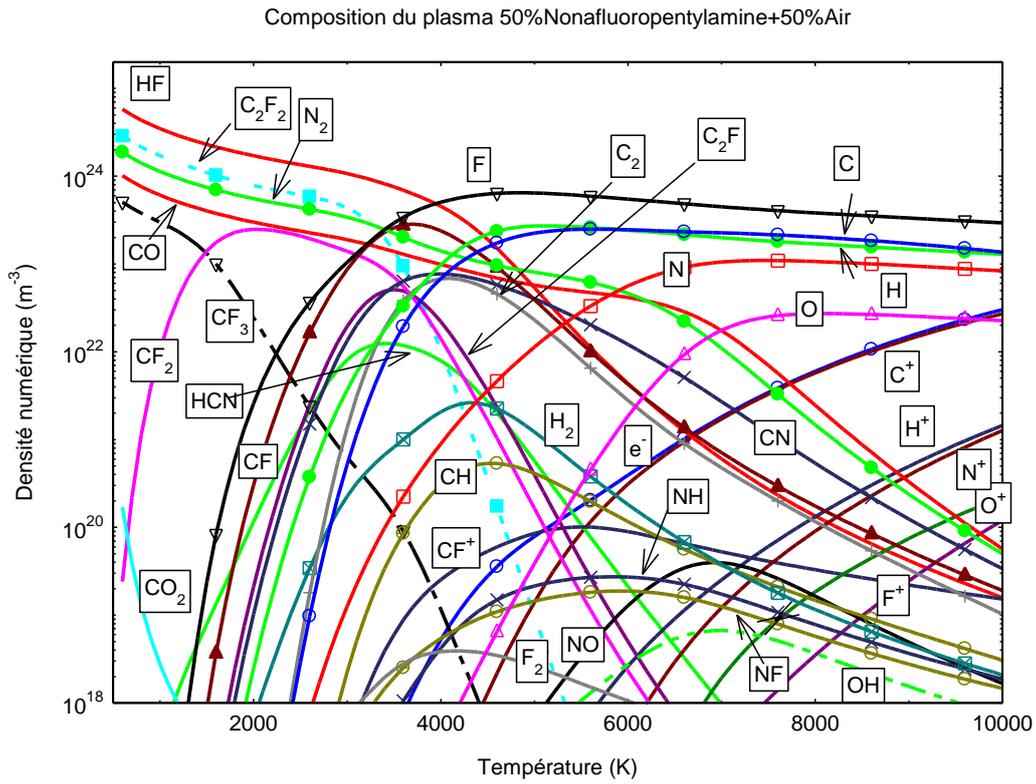

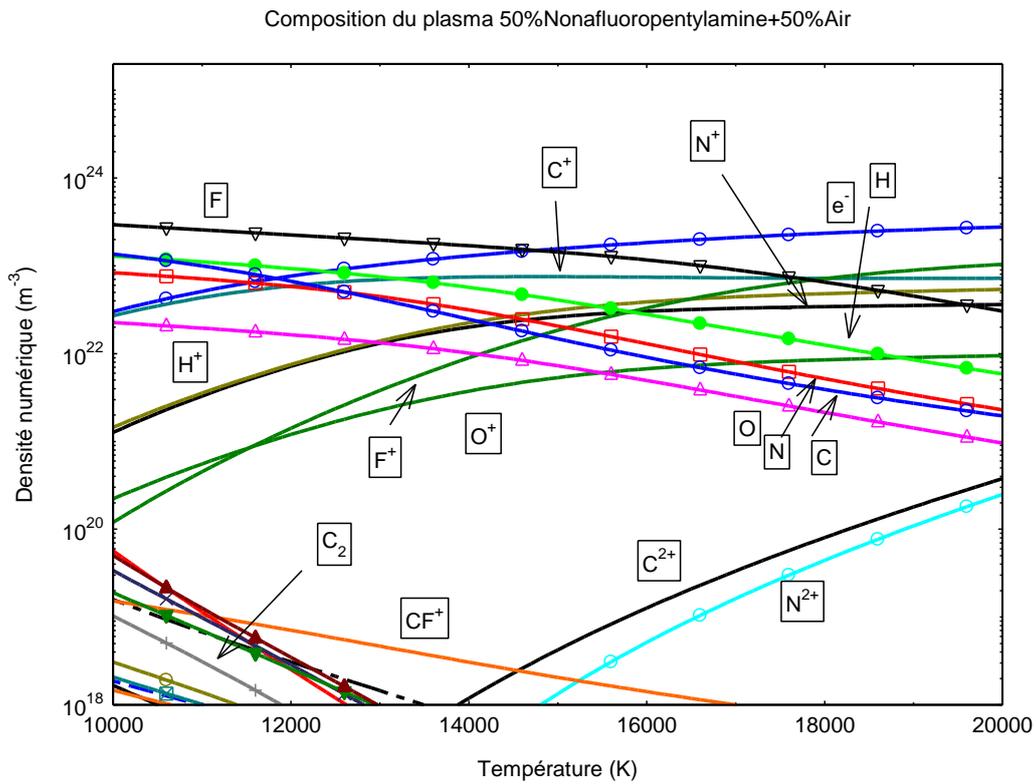

**Figure 7 :** Évolutions en fonction de la température des densités numériques des espèces chimiques du plasma composé de 50% de nonafluoropentylamine et 50% d'air en volume à la pression atmosphérique.

Ainsi, nous observons l'apparition d'espèces chimiques nocives pour l'environnement et pour l'homme : $CH_4$, $CF_2$, $CO_2, CO, NO, HCN, HF$.

Le dioxyde de carbone ($CO_2$) est un gaz peu toxique à faible dose mais il est l'un des principaux gaz à effet de serre qui mène au changement climatique. À très forte concentration, il peut cependant provoquer des malaises et des maux de tête [25]. Le monoxyde de carbone ($CO$) se fixe sur l'hémoglobine du sang, avec une affinité 200 fois supérieure à celle de l'oxygène, ce qui peut influencer le cerveau et le cœur. L'inhalation de $CO$ entraîne des maux de tête et des vertiges et peut mener à la mort [25]. Le monoxyde d'azote (*NO*), est un gaz incolore qui au contact de l'air se transforme en un gaz toxique roux qui le dioxyde d'azote (*NO₂*). Le *NO₂* provoque une hyperréactivité bronchique chez les asthmatiques et se transforme dans l'atmosphère en acide nitrique [25]. Le méthane (*CH₄*) est un puissant gaz à effet de serre qui contribue au réchauffement climatique. Le (*CH₄*) a un impact sur l'effet de serre environ vingt-et-une fois plus puissant que le dioxyde de carbone (*CO₂*) [26]. Le cyanure d'hydrogène (*HCN*) est très toxique quelle que soit la voie d'exposition. Le fluorure d'hydrogène (*HF*) est un gaz incolore à odeur piquante. C'est un gaz hautement toxique par inhalation, ingestion ou contact cutané. Il réagit avec l'eau pour former de l'acide fluorhydrique, très corrosif et extrêmement dangereux pour l'environnement (modification du pH de l'eau).

### 4.4. Influence du pourcentage d'air sur les densités des espèces chimiques ($HCN, HF, CF_2, CO$).

Pour observer l'influence du pourcentage d'air sur les compositions d'équilibre des plasmas de mélange, nous avons représenté sur les figures 8 et 9 l'évolution en fonction de la température des densités numériques de certaines espèces chimiques ($HCN, HF, CF_2, CO$) nocives pour l'environnement et l'homme pour différentes valeurs du pourcentage d'air dans les mélanges. On remarque que l'influence du taux d'air sur les densités numériques des espèces chimiques du mélange de trifluoroéthylamine est similaire à son influence dans le cas du mélange de nonafluoropentylamine. Il ressort de cette étude que, l'apport de l'air augmente les concentrations de ces espèces chimiques ($HCN, CF_2, CO$), tandis que celle ($HF$) diminue légèrement dans le plasma pour des températures inférieures à 15.000 K. Par conséquent, pour éliminer les déchets tout en réduisant considérablement les risques pour l'homme et pour l'environnement, la destruction des molécules nocives doit être réalisée à des températures supérieures à 15.000 K.

Dans le cas du plasma de mélange de nonafluoropentylamine et d'air, nous observons une évolution analogue des densités numériques des espèces chimiques $HCN, HF, CF_2$ et $CO$ en fonction de la température pour également différentes valeurs du pourcentage d'air.

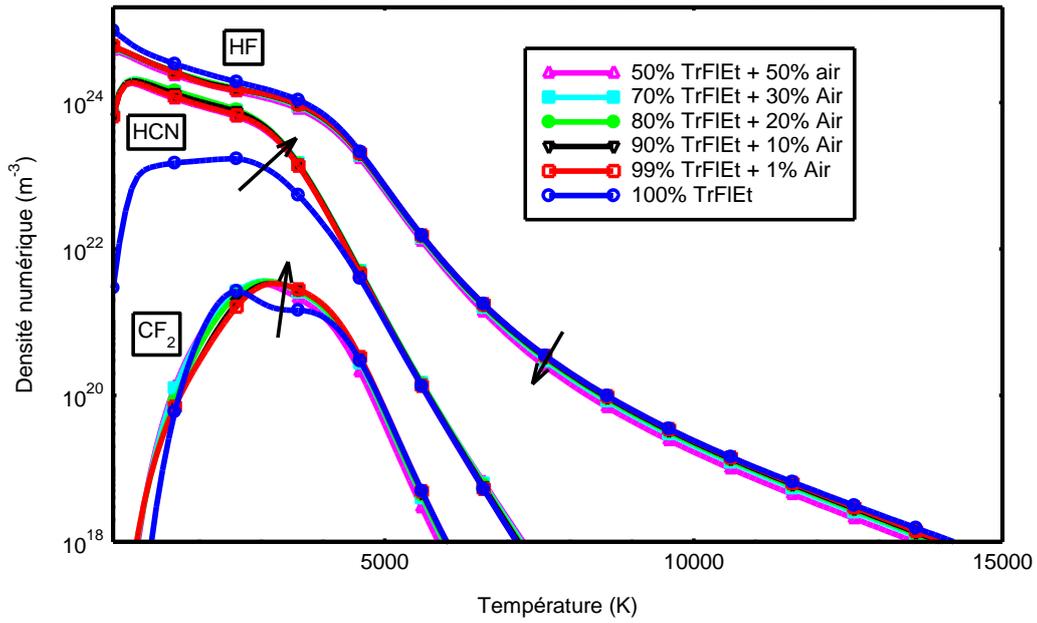

**Figure 8 :** Influence du pourcentage d'air en volume sur les densités des espèces chimiques (*CF$_2$, HCN, HF*) du plasma de trifluoroéthylamine

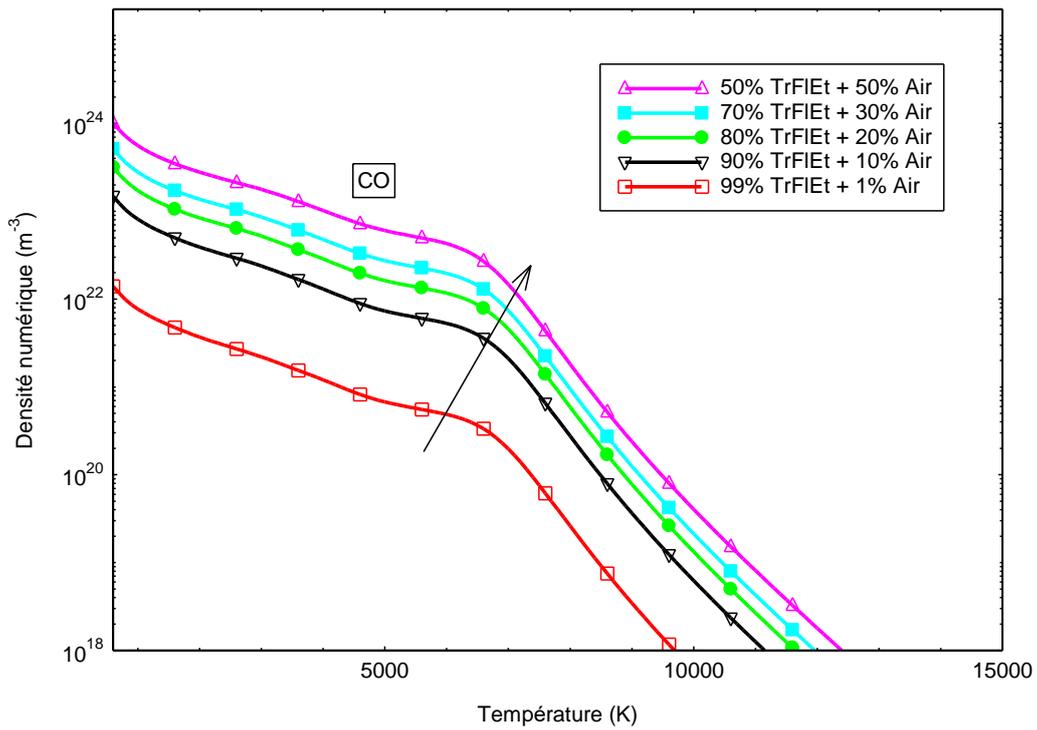

**Figure 9 :** Influence du pourcentage d'air en volume sur les densités de monoxyde de carbone (CO) du plasma de trifluoroéthylamine

**Conclusion**

Dans cet article, nous avons présenté les résultats des calculs de composition d'équilibre et leurs interprétations. La composition chimique des plasmas de mélanges d'air, de trifluoroéthylamine et de nonafluoropentylamine a été déterminée à l'équilibre thermodynamique local et à la pression atmosphérique. Nous avons également donné les compositions d'équilibre des plasmas de mélanges de trifluoroéthylamine-air et de nonafluoropentylamine-air. Il ressort des résultats de calcul de la composition des plasmas que l'apport d'air dans le mélange augmente les concentrations des espèces chimiques comme $CF_2, CO, HCN$ aussi bien dans le cas du trifluoroéthylamine que celui du nonafluoropentylamine tandis que celle de HF diminue légèrement. Ces espèces chimiques n'existent dans le plasma que pour des températures inférieures à 15.000 K (T < 15.000 K). Ces espèces chimiques dangereuses peuvent donc apparaitre à basse température lors du refroidissement du plasma puis du gaz. Il est donc nécessaire d'effectuer un lavage des fumées avant de les relâcher dans l'atmosphère.

L'incinération de ce nouveau type de déchets composés de nouvelles molécules basées sur des fluoroalkylamines est complexe. En effet, des molécules à fort impact écologique et toxiques pour l'homme sont formées. Par conséquent, il est aussi nécessaire de mettre en place une véritable politique de gestion des déchets afin d'éviter d'incinérer par erreurs des déchets contenant des fluoroalkylAmines.